\documentclass[preprint,12pt]{elsarticle}

\usepackage{amssymb}
\usepackage[margin=2.5cm]{geometry} 
\usepackage{lipsum}
\usepackage{graphicx}
\usepackage{caption}
\usepackage{subcaption}
\usepackage{diagbox}
\usepackage{multirow}
\usepackage{booktabs}
\usepackage{diagbox}
\usepackage{amsmath}
\usepackage{url}
\usepackage{enumitem}
\usepackage{xurl}
\usepackage{multirow} 

\usepackage[usenames,dvipsnames]{color}
\usepackage{multirow}
\usepackage{amssymb}
\usepackage{amsmath}
\usepackage{graphicx}
\usepackage{float} 
\usepackage{natbib}

\setcitestyle{square}

\makeatletter
 \def\ps@pprintTitle{%
  \let\@oddhead\@empty
  \let\@evenhead\@empty
  \def\@oddfoot{}%
  \let\@evenfoot\@oddfoot}
 \makeatother
\begin{document}
\begin{frontmatter}



\title{Assessing Workers’ Neuro-physiological Stress Responses to Augmented Reality Safety Warnings in Immersive Virtual \\Roadway Work Zones}

\author[inst1]{Fatemeh Banani Ardecani}
\author[inst1]{Omidreza Shoghli\corref{cor1}}
\affiliation[inst1]{organization={William State Lee College of Engineering, University of North Carolina at Charlotte},
            addressline={9201 University City Blvd}, 
            city={Charlotte},
            postcode={28223}, 
            state={North Carolina},
            country={USA}}

\cortext[cor1]{Corresponding author: oshoghli@charlotte.edu}

\begin{abstract} 
This paper presents a multi-stage experimental framework that integrates immersive Virtual Reality (VR) simulations, wearable sensors, and advanced signal processing to investigate construction workers’ neuro-physiological stress responses to multi-sensory AR-enabled warnings. Participants performed light- and moderate-intensity roadway maintenance tasks within a high-fidelity VR roadway work zone, while key stress markers of electrodermal activity (EDA), heart rate variability (HRV), and electroencephalography (EEG) were continuously measured. Statistical analyses revealed that task intensity significantly influenced physiological and neurological stress indicators. Moderate-intensity tasks elicited greater autonomic arousal, evidenced by elevated heart rate measures (mean-HR, std-HR, max-HR) and stronger electrodermal responses, while EEG data indicated distinct stress-related alpha suppression and beta enhancement. Feature-importance analysis further identified mean EDR and short-term HR metrics as discriminative for classifying task intensity. Correlation results highlighted a temporal lag between immediate neural changes and subsequent physiological stress reactions, emphasizing the interplay between cognition and autonomic regulation during hazardous tasks. 
\end{abstract}


\end{frontmatter}

\section{Introduction}
\label{sec1}
Roadway work zones are indispensable for maintaining and upgrading transportation infrastructure, yet they also pose significant safety hazards. Shifts in traffic patterns, reduced lane widths, and ongoing construction and maintenance activities increase the risk of crashes, injuries, and fatalities. Moreover, these incidents often lead to major traffic delays due to constrained maneuvering space within work zones \cite{FHWA_Work_Zone_Facts_Stats}. 
It was estimated that roadway work zones have experienced an average of 125 worker fatalities per year over the past decade \cite{WZSIC2025Data, zhang2025improving}. Beyond the human toll, the economic burden of these incidents has grown, with the lifetime comprehensive cost per fatality reaching \$13.2 million in 2023 \cite{USDOT2023VSL}. These figures emphasize that despite the implementation of standardized traffic control measures, there is a need to enhance existing safety strategies and explore innovative technologies to better protect workers and improve efficiency in roadway work zones. Recent advances in Artificial Intelligence, the Internet of Things (IoT), and real-time communication architectures further underscore the feasibility of transitioning from reactive to proactive safety systems \cite{sabeti2022toward}.

Augmented Reality (AR) has emerged as a transformative technology with the potential to enhance situational awareness and improve safety in high-risk environments. Recent advancements in both hardware and software have significantly improved AR's effectiveness in delivering real-time warnings within the transportation and construction sectors \cite{sabeti2021toward,ramos2022proposal,li2018critical,gilson2020leveraging}. In roadway work zones, where workers face dynamic and cognitively demanding conditions, AR-assisted multi-sensory warnings offer a promising approach to mitigating risks. Despite the potential benefits of AR-assisted warnings in alerting workers to dangers and reducing their reaction time, roadway work zones remain highly stressful environments due to constant exposure to hazards, loud noises, and fast-moving traffic. Therefore, their effects on stress must be carefully assessed to ensure they enhance safety without adding cognitive strain. Despite AR's promising potential, significant gaps in knowledge persist including (1) insufficient understanding of physiological stress responses to multi-sensory AR-enabled warnings in roadway work zones, (2) limited evidence on the impact of varying task intensities on physiological stress reactions, and
(3) lack of research exploring the temporal dynamics between cognitive neural indicators of stress and subsequent physiological responses triggered by AR warnings.

To address these gaps in knowledge, the objectives of this study are to 
(1) Evaluate the effect of task intensity on physiological stress responses of workers exposed to multi-sensory AR-enabled warnings as measured by HR, HRV and EDR metrics. 
(2) Rank the key stress features among HR, HRV and EDR metrics to distinguish between low- versus moderate-intensity tasks when responding to multi-sensory AR-enabled safety warnings.
(3) Analyze the effect of task intensity on workers' neurological stress responses to multi-sensory AR-enabled warnings, as measured by cognitive metrics. 
(4) Determine the temporal relationship between neural indicators of stress and subsequent physiological responses following multi-sensory AR-enabled warnings.

This study advances the body of knowledge on stress assessment in roadway construction workers by examining their physiological responses to multi-modal AR warning system as well as investigating the temporal relationship between neural indicators of stress and subsequent physiological responses following multi-sensory AR-enabled warnings. By analyzing EEG-derived cognitive metrics alongside physiological markers such as heart rate (HR), heart rate variability (HRV) and electrodermal response (EDR), this research provides deeper insights into the mechanisms underlying workers' stress responses to multi-modal AR warning systems. Additionally, it identifies and ranks key stress features across HR, HRV and EDR to differentiate between low- and moderate-intensity tasks, offering a systematic approach to evaluating workload and cognitive strain in real-world construction settings.

Beyond its contributions to the body of knowledge, this study has practical implications for enhancing worker safety and optimizing AR applications in professional environments. By establishing reliable stress assessment metrics, it lays the groundwork for the development of real-time monitoring systems capable of detecting and mitigating stress with minimal data processing. These findings can inform the design of adaptive AR safety interventions tailored to individual stress thresholds, ultimately improving worker performance and reducing accident risks. Furthermore, the proposed methodology serves as a scalable framework that can be adapted to other high-risk industries, such as manufacturing and military operations, where AR technology holds significant potential for improving situational awareness and safety.

\section{Related Works}
\subsection{Physiological Measures of Stress in Occupational Safety}
Stress is defined by sudden changes in the environment that require the human body to react, respond, and adapt accordingly \cite{asif2019human, cohen1997measuring}. 
Among physiological indicators, heart rate (HR) and heart rate variability (HRV) are widely utilized, with HRV reflecting fluctuations in the time intervals between successive heartbeats (inter-beat intervals, IBI) \cite{shaffer2017overview}. The analysis of HRV is commonly conducted in both the time and frequency domains, where time-domain measures quantify variations in IBI over a given period \cite{stuyck2022validity}. One frequently used time-domain metric is pNN50, which represents the proportion of successive NN intervals that differ by more than 50 milliseconds. This measure was first introduced in a foundational study by Ewing et al. (1984), where the NN50 count was proposed as a marker of autonomic function \cite{ewing1984new, costa2014interoceptive}. In the frequency domain, HRV metrics are determined by analyzing the distribution of signal power across specific frequency bands. These bands include low frequency (LF; 0.04–0.15 Hz), and high frequency (HF; 0.15–0.4 Hz). The relative or normalized power in these bands provides insights into autonomic nervous system activity \cite{shaffer2017overview}.

In addition to HR and HRV, electrodermal activity has also been used as a physiological indicator of stress-related autonomic responses. EDA refers to fluctuations in the skin's electrical properties resulting from sweat gland activity, which is regulated by the sympathetic nervous system. When an individual experiences physical exertion or mental stress, sympathetic activation increases, leading to greater sweat gland activity and, consequently, higher skin conductance. As a result, EDA has been employed as a measure of sympathetic nervous system activation in response to stress \cite{setz2009discriminating, picard2016multiple}.

\subsection{EEG Correlates of Stress}
Recent neuroscience research has underscored the utility of EEG for assessing stress, given the robust correlation between psychological stress and variations in EEG power \cite{seo2010stress}. In particular, stress-related changes are prominently observed in the frontal lobe \cite{konar2023novel}. Increases in beta power,  a phenomenon especially notable which plays a critical role in executive functions such as decision-making, working memory, and attention, have been associated with elevated stress \cite{ arnsten2009stress, moghaddam2004effect, starcke2012decision}. During stressful events, the brain typically ramps up its overall activity \cite{Lee10, wheelock2016prefrontal}, with the heightened beta activity reflecting increased alertness, focus, and concentration \cite{engel2010beta, vanhollebeke2022neural}. For instance,  Seo et al. \cite{seo2010stress} conducted a resting-state experiment comparing high- and low-stress individuals based on self-reported daily hassles. EEG analysis revealed elevated beta power in the frontal and temporal regions among high-stress participants, indicating heightened arousal and supporting beta activity as a marker of chronic stress. Similarly, Jena et al. \cite{jena2015examination} monitored EEG activity among medical students during normal conditions and under examination stress. A significant increase in beta wave activity (12–30 Hz) was observed during stress, especially in students with mild to moderate baseline stress, indicating heightened alertness and cognitive strain under academic pressure. Alongside increased beta activity, a decline in alpha wave power is commonly observed; since elevated alpha activity is generally linked to relaxation and calmness \cite{8758154, vanhollebeke2022neural}, its suppression may signal a reallocation of neural resources toward processing immediate demands or potential threats \cite{braboszcz2011lost, reisman1997measurement}. Consequently, while alpha wave activity has traditionally received significant attention, recent research increasingly acknowledges beta activity as a complementary indicator of the neural response to stress. Supporting this perspective, Waili et al. \cite{TuerxunWaili_2020} used  a low-cost, wearable EEG device to monitor brain activity during calm and stress-inducing video scenarios, finding that participants under stress consistently exhibited alpha suppression and beta enhancement. This study validates the neural markers of stress and demonstrates that even single-channel EEG systems can effectively detect stress-related brainwave changes in controlled environments.

\subsection{Quantifying Physiological and Neurological Metrics in AR-Assisted Environments}
In recent years, advancements in wearable sensor technology have enabled more accurate, objective, and continuous monitoring of neuro-physiological responses. Consequently, there has been increased research interest in applying these sensors to evaluate physiological markers across various domains, including challenging and high-risk environments such as construction, which is both physically demanding and dynamic. Given the high physical and cognitive demands of construction work, understanding and analyzing workers’ stress responses has drawn considerable attention, as stress significantly impacts workers’ performance, health, and overall job efficiency. Indeed, the Chartered Institute of Building reported that 68\% of construction workers suffer from excessive stress due to the demanding nature of their work \cite{campbell2006occupational}.

Previous work has utilized physiological measurements such as electrodermal activity (EDA) \cite{newton2024measuring}, photoplethysmography (PPG) \cite{ojha2024quantifying}, and electroencephalography (EEG) \cite{jebelli2018eeg} to examine chronic stress in construction contexts, focusing on how variations in task complexity, workload, and physical exertion affect workers’ physiological responses \cite{jebelli2019applicationI, jebelli2019applicationII}. While investigations into the relationship between job intensity and stress have expanded, research remains limited on how AR technologies modulate these same physiological responses. A small number of recent contributions highlight the growing interest in understanding AR’s impact on users’ stress levels and cognitive workload. For example, Kia et al. \cite{kia2021effects} examined participants’ AR-based performance using functional near-infrared spectroscopy (fNIRS) for cerebral oxygenation, Short Stress State Questionnaire for stress, NASA-TLX for subjective workload, and overall task performance. Their AR scenario encompassed an omni-directional pointing task, using gaze and pinch gestures to select virtual targets, and a cube placing task in 3D space. Along similar lines, Kim et al. \cite{kim2020evaluation} characterized biomechanical stress (neck and shoulder), self-reported discomfort, and usability when participants moved and manipulated virtual cubes at varying distances and sizes in AR; their study also included a web-browsing task requiring gaze and pinch interactions with on-screen links.

Beyond these select examples, a body of work emphasizes the varying ways AR can affect physiological and neurological functioning. Recent research targeting physiological measures has leveraged ECG-based heart-rate variability (HRV) \cite{tsai2018effect, alessa2023evaluating}, photoplethysmography-based HRV, and EDA-based electrodermal responses \cite{dey2022effects, brunzini2022comprehensive, nardelli2020recognizing}, capturing how AR tasks can increase emotional arousal and perceived cognitive demands. Studies focusing on neurological metrics, primarily EEG \cite{krugliak2022towards, vortmann2019eeg, giannopulu2022synchronised, atici2021effects}, report shifts in power spectra (e.g., low-frequency posterior alpha for visual cognition or frontal theta for attentional processes) depending on the complexity of AR-based interactions. Finally, neuro-physiological research integrating both peripheral (HRV, EDA) and cortical (EEG) measures has shown that AR guidance and training can produce quantifiable changes in stress and workload markers \cite{callara2024behavioral, jarvela2021augmented, eom2023investigation}, emphasizing how combined data streams offer deeper insight into user responses.  However, the influence of AR safety warnings on users’ stress and cognitive workload, particularly in high-risk settings such as construction, is underexplored. This gap highlights the need for further investigation into how AR-based interventions might enhance both safety and user well-being in physically and cognitively demanding work environments.

\section{Method}
\begin{figure}
    \centering
    \includegraphics[width=.98\linewidth]{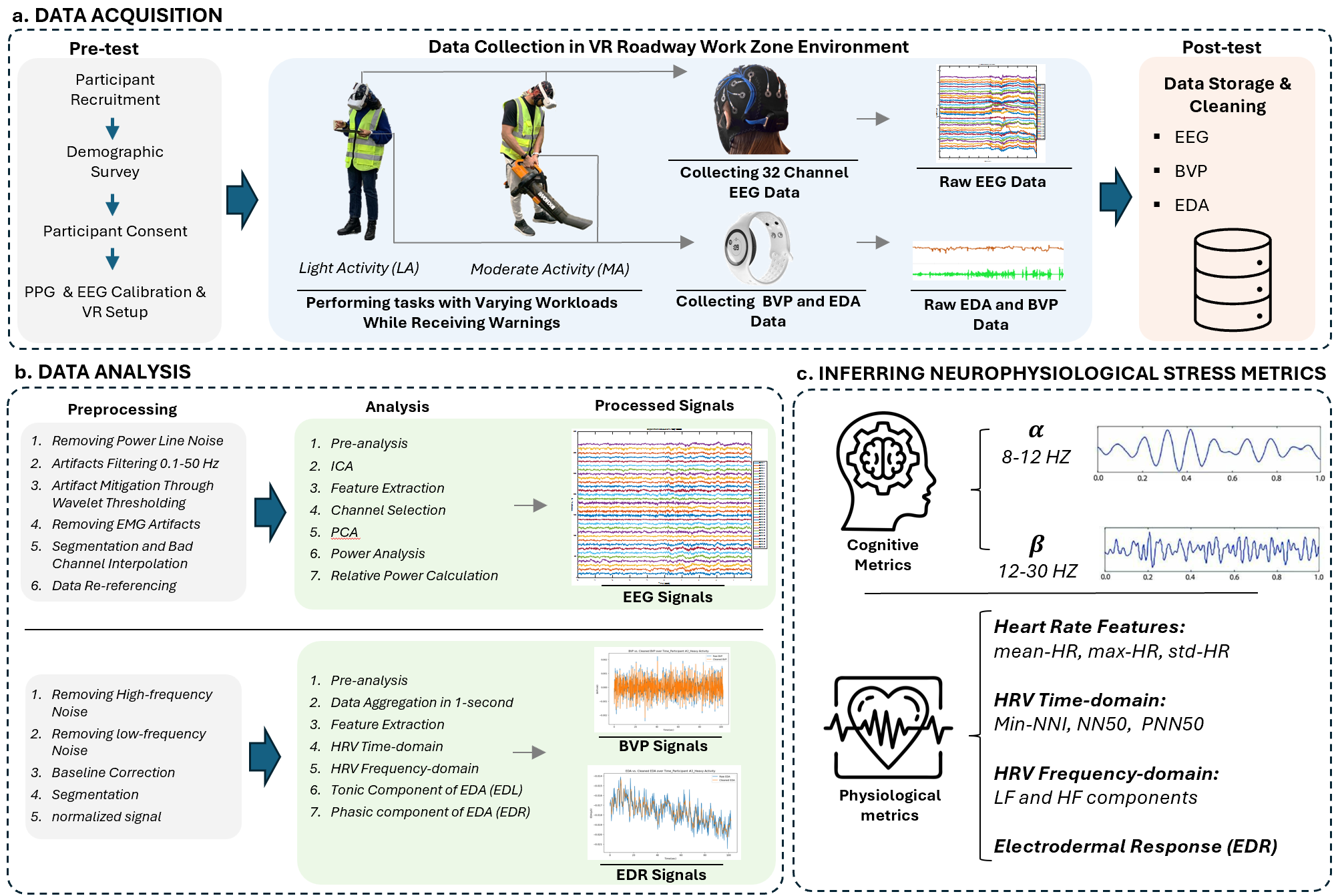}
    \captionsetup{skip=8pt}
    \caption{Overview of the primary methodological components: (a) Data acquisition, (b) Data analysis, (c) Inferring neuro-physiological stress metrics}
    \label{EEG_PPG_method_Overview}
\end{figure}

To assess the neuro-physiological stress responses of construction workers exposed to multi-sensory AR warnings, a multi-stage methodology was developed. The proposed methodology integrates immersive virtual reality simulations, real-time data acquisition, and detailed signal analyses to infer cognitive metrics, as depicted in Figure~\ref{EEG_PPG_method_Overview}. This study was approved by the Institutional Review Board (IRB) at the University of North Carolina at Charlotte (21-0357). Participants provided informed consent after being briefed on the study objectives and procedures. Each task was designed for completion within two minutes, and task order was randomized to minimize bias and learning effects.

\label{sec3}

\subsection{Experimental Design and Procedure}
\label{sec3-1}
To achieve the study’s objectives, a high-fidelity virtual reality (VR) environment was developed to simulate a highway work zone, incorporating the functionality of AR glasses. This environment was designed to replicate a short-duration roadway work zone, a temporary setup lasting up to one hour for routine maintenance and operational tasks based on the guidelines from the Manual on Uniform Traffic Control Devices (MUTCD) \citep{FHWA2009MUTCD}. Within this environment, participants performed two routine roadway maintenance tasks that effectively replicated real-world activities. These tasks, categorized as light and moderate intensity, were selected to represent different levels of physiological demand commonly faced in roadway maintenance settings. This classification aligns with standard methodologies in research examining fatigue, stress, and physiological and cognitive factors, ensuring that findings reflect real-world construction scenarios under varying conditions \citep{jebelli2019applicationII, chen2016revealing, wang2019detecting}.

\begin{figure}[H]
    \centering
    \includegraphics[width=0.85\linewidth]{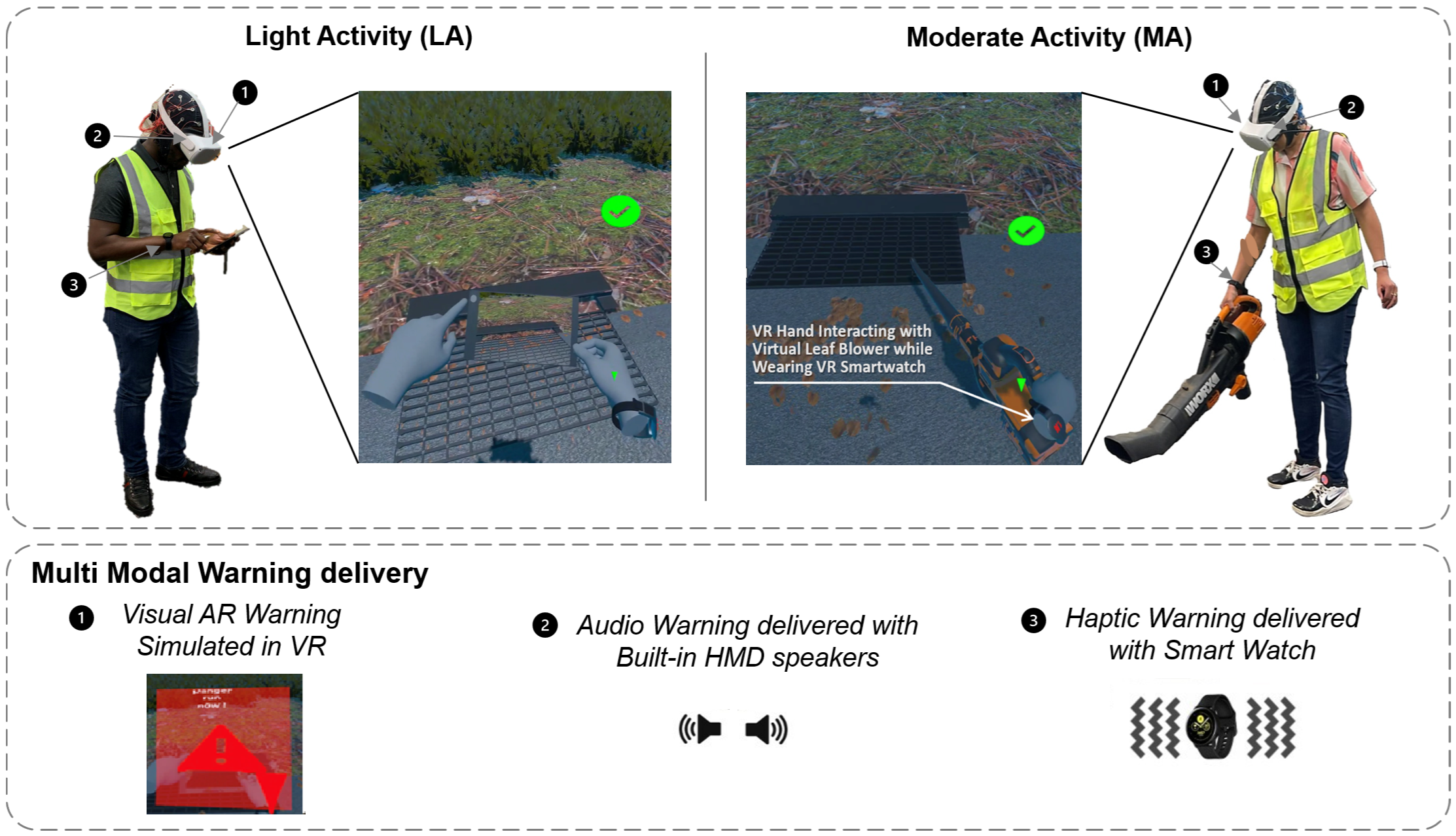}
    \captionsetup{skip=5pt}
    \caption{Experimental setup of participants performing light and moderate intensity tasks in the immersive virtual reality environment of roadway work zones, and multi modal safety warning delivery mechanisms}

    \label{ExperimentTaskDesign}
\end{figure}

\textit{Light Activity (LA)} involved inspecting an obstructed drop inlet using a tablet and documenting the defect by capturing images, as illustrated in Figure~\ref{ExperimentTaskDesign}. Tasks involving minimal movement, such as inspections and documentation, typically impose low physical and cognitive demands \citep{jebelli2019applicationII}. Participants physically held a tablet while interacting with its virtual counterpart in the VR environment. The VR simulation also included auditory feedback for enhanced realism.
Tasks involving minimal movement, such as briefing, reading construction plans, and inspections, are typically classified as low-intensity activities due to their limited physical and cognitive demands \citep{jebelli2019applicationII}. In developing the LA task in the VR simulation, a virtual model of a tablet with image-capturing capability was created within the 3D environment. Participants held a physical tablet in the real world while viewing its virtual counterpart in the VR environment. Additionally, a virtual button for capturing images was incorporated into the virtual model, complete with a sound effect to enhance the user experience.

\textit{Moderate Activity (MA)} involved clearing debris from a drop inlet using a leaf blower, representing tasks that require greater physical effort. In construction, moderate-intensity activities include clearing the work zone, handling tools, transporting small materials, and performing measurements or minor modifications to materials \citep{jebelli2019applicationII}. Medium-level tasks, such as site cleaning with light tools, are often associated with both physical and cognitive demands, making them particularly relevant for studying workload variations \citep{jebelli2018eeg}. In this task, a virtual representation of the same real-life leaf blower that participants physically held in the VR simulation was used. To enhance realism, the audio effect of the leaf blower was replicated in the VR environment, and collision detection capabilities were incorporated to simulate the removal of obstructions from the drop inlet. 

Figure \ref{ExperimentTaskDesign} illustrates examples of these interactions and the developed activities.  
The experimental procedure involved five key steps: Participants
(1) navigated to the work vehicle within the immersive VR environment;
(2) wore simulated AR glasses integrated into the VR environment;
(3) experienced multimodal safety alerts delivered through the virtual AR glasses and a physical smartwatch;
(4) performed a light-intensity inspection task by capturing images using a tablet; and
(5) carried out a moderate-intensity maintenance task, clearing debris with a leaf blower while simultaneously responding to safety warnings.

During task performance, participants received multimodal safety warnings delivered through three sensory channels (visual, haptic, and auditory). These warnings were developed according to our previously established framework and warning desing \citep{sabeti2024augmented, sabeti2024mixed}. The multimodal warning included three simultaneous cues as shown in Figure \ref{ExperimentTaskDesign}: 
(1) Visual warning appeared on a simulated AR display within the VR environment
(2) Auditory warning, a brief, high-pitched beep at 44,100 Hz for 0.2 milliseconds, was broadcast through the VR headset’s built-in speakers 
(3) Haptic warning was delivered using a smartwatch haptic feedback 

Throughout the experiment, physiological responses and brain activity were continuously recorded using wearable devices, including the Empatica Embrace Plus and the Emotiv FLEX 2 Gel. The Empatica Embrace Plus device includes embedded sensors for PPG and EDA to measure physiological responses, along with additional sensors such as an accelerometer, gyroscope, and temperature sensor for contextual movement and environmental data. The Emotiv FLEX 2 Gel is a gel-based EEG device with 32 channels placed according to the standard 10–20 system. These devices enabled real-time capturing of cognitive and physiological states in a simulated roadway work zone environment. The experimental design followed a within-subject framework, where each participant served as their own control. To establish a reference for comparison, PPG and EEG data were recorded prior to the activation of AR warnings, serving as a baseline. This methodological choice was based on three key principles. First, it enabled a direct comparison of neuro-physiological responses in the presence and absence of multi-sensory warnings, reducing variability across individuals. Second, baseline measurements were taken during task engagement rather than rest periods, ensuring that inherent differences between light and moderate tasks were accounted for. Using rest as a baseline would not have captured the task-specific neural activity necessary for evaluating stimulus-related effects \cite{brauns2021head, tarailis2024functional}. Third, this approach aligns with established practices in neurophysiology research. Prior studies emphasize the significance of pre-stimulus activity as a baseline for isolating responses to experimental stimuli \cite{luck2014introduction, cohen2014analyzing}, supporting the reliability of this method in detecting changes induced by AR warnings.

\subsection{Experimental Equipment}
\label{sec3-2}
The immersive VR environment used in the study was developed with the Unity 3D game engine \citep{unity} and was delivered via the Oculus Quest 2 headset \citep{quest2}. Within this setup, augmented reality warnings were simulated through a virtual AR display, allowing participants to experience multi-sensory safety alerts. Physical and virtual tools were integrated into the environment to ensure realistic task execution. Participants in the Light Activity (LA) condition utilized an Amazon Fire Tablet (Fire HD 10, Fire OS 7) \citep{Amazon2024}, while those in the Moderate Activity (MA) condition operated a WORX WG509 12 Amp leaf blower \citep{WORX2023}. To enhance immersion, digital replicas of these tools were also incorporated into the VR simulation. The haptic feedback component of the warnings was delivered using a Samsung Galaxy Smartwatch Active (SM-R500NZKAXAR) \citep{SamsungElectronicsAmerica2024}, controlled via the Tizen Native framework \citep{tizen}. A predefined vibration pattern from the smartwatch's API provided consistent tactile alerts throughout the experiment.

\begin{figure}[H]
    \centering
    \includegraphics[width=0.8\textwidth]{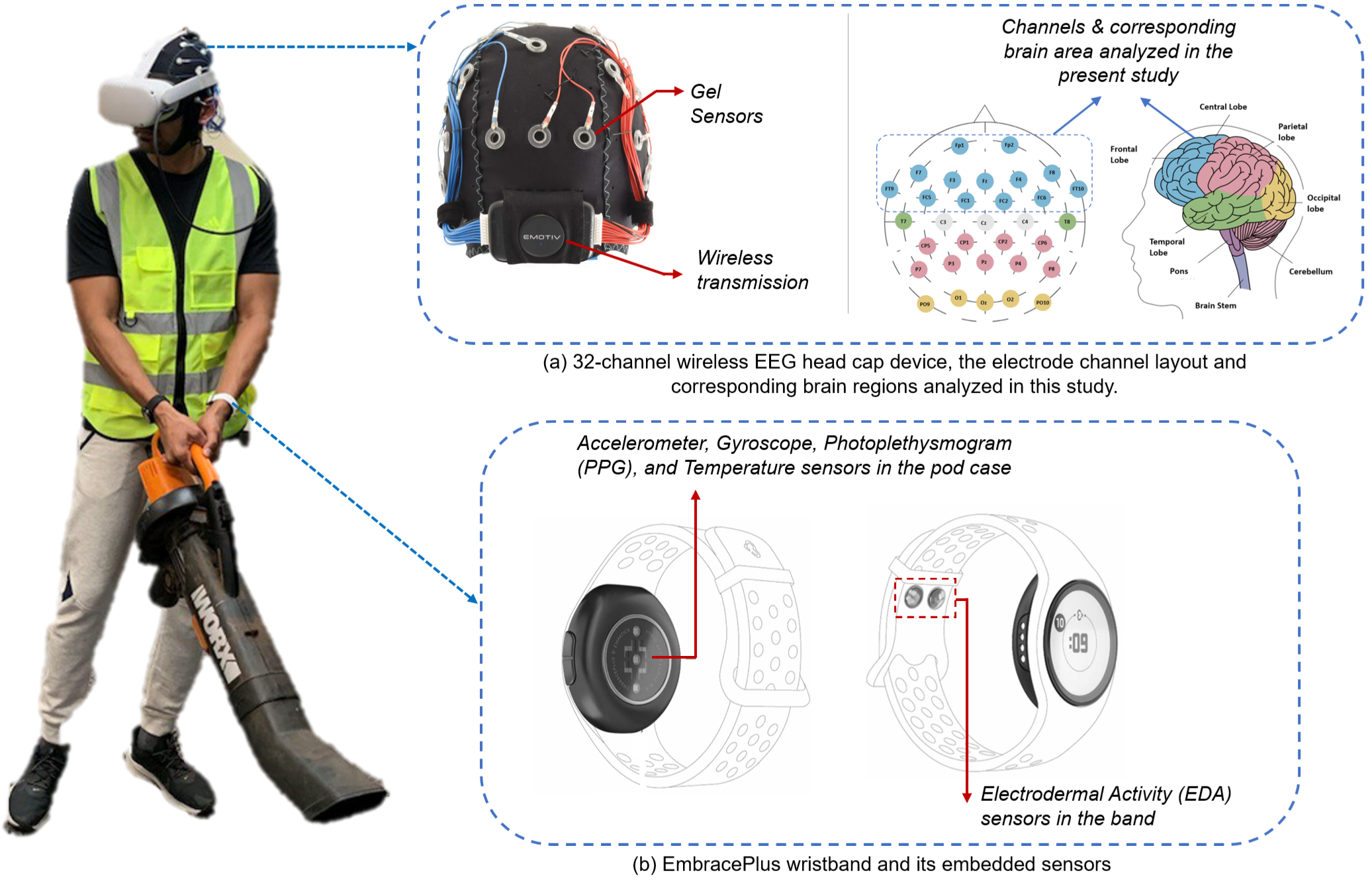}
    \captionsetup{skip=10pt}
    \caption{(a) Utilized EEG device, spatial configuration of EEG electrodes, and targeted brain areas  (b) Utilized wearable wristband and its embedded sensors 
    }
    \label{Cerebral-Cortex-and-physiological}
\end{figure}

To monitor physiological responses in the virtual roadway work zone, a wrist-worn sensing device was employed. The Embrace Plus wearable wristband \citep{empatica} was utilized to collect PPG and EDA signals, offering a comprehensive assessment of participants' physiological states, as shown in Figure \ref{Cerebral-Cortex-and-physiological}(a). Data were recorded at the device's highest available sampling rates, with PPG captured at 64 Hz and EDA and ST at 4 Hz.

Neurological data were captured using the Emotiv Flex 2 EEG system \citep{EMOTIV2024}, a wireless 32-channel headcap designed for high-resolution brain activity analysis. The system consists of 32 active electrodes and two reference electrodes, positioned in accordance with the 10-20 international electrode placement system. To ensure high-quality signal acquisition, gel-based sensors were used, optimizing conductivity while minimizing impedance. EEG data were sampled at 128 Hz and wirelessly transmitted via Bluetooth to a host computer for real-time processing and storage. Figure \ref{Cerebral-Cortex-and-physiological}(b) provides an overview of the EEG electrode placements used in this study.

\subsection{Participants}
\label{sec3-3}
This study initially recruited 20 participants with the goal of detecting large effect sizes (Cohen’s d = 0.80) at a median statistical power of 0.7 and a significance level of 0.05. However, due to significant data loss, physiological signals from three participants were incomplete, reducing the final dataset to 17 participants. A post-hoc power analysis confirmed that this sample size remains sufficient for detecting large effect sizes (Cohen’s d = 0.80) with a median power of 0.7, though the revised significance level is estimated to be 0.079. 
The final sample consisted of 13 males and 4 females, with an average age of 27.12 years (SD = 6.82). Among them, 12 participants had an average work experience of 3.92 years (SD = 4.78), while the remaining five had no prior professional experience in similar tasks.

\subsection{Physiological Data Processing and Analysis}
\label{sec3-4}
Physiological metrics, including HR, HRV and EDA, were used to assess autonomic nervous system responses. HR and HRV features were extracted from  blood volume changes in peripheral tissue using light absorption by the PPG sensor, while EDA signals were processed to evaluate sympathetic arousal. 
Given the importance of capturing transient autonomic responses, the analysis focused on a 10-second period following the second warning, aligning with prior research suggesting that ultra-short intervals, such as 10-second segments, can provide meaningful estimations of physiological metrics \cite{stuyck2022validity}. This approach enables the assessment of rapid changes in electrodermal and cardiovascular activity, which are critical for understanding stress responses and attentional shifts in dynamic environments.

\subsubsection{Heart Rate (HR) and Heart Rate Variability (HRV)}
\label{sec3-4-1}
HR and HRV features were extracted from the Interbeat Interval (IBI) data, which represents the time interval between successive heartbeats. The IBI data was derived from the Blood Volume Pulse (BVP) signal captured by the PPG sensor \cite{lohani2019review}. Since BVP signals exhibit a strong correlation with IBI measurements, the wrist-worn device employs a proprietary algorithm to process the raw BVP data, mitigating artifacts and erroneous peaks introduced by motion noise or environmental disturbances \cite{stuyck2022validity}. 
This refined IBI data serves as the basis for analyzing HR features—including maximum (Max-HR), mean (Mean-HR), and standard deviation (Std-HR)—as well as HRV features, encompassing both time-domain and frequency-domain metrics to assess autonomic nervous system activity.

\textit{Time-domain features} extracted from the IBI data included the minimum normal-to-normal interval (min-NNI), standard deviation of consecutive  NNI (SDNN), the root mean square of successive NNI differences (RMSSD), the number of successive NN intervals differing by more than 50 ms (NN50), and the percentage of NN50 relative to total NN intervals (PNN50). These parameters reflect autonomic nervous system activity and physiological responses to stress \cite{kim2018stress}.

\textit{Frequency-domain} involved decomposing HRV signals into low-frequency (LF: 0.04–0.15 Hz) and high-frequency (HF: 0.15–0.4 Hz) components, which represent sympathetic (SNS) and parasympathetic (PNS) nervous system influences, respectively. While the SNS is responsible for rapid responses to external stimuli, the PNS facilitates relaxation and recovery after a reaction \cite{glick1965relative}.

To extract these features, the pyHRV package \cite{gomes2019pyhrv}, an open-source Python toolkit, was utilized. This process involved automated HR and HRV computation from IBI data, facilitating the analysis of statistical parameters from a series of HR as well as  both time-domain and frequency-domain features of HRV. The preprocessing stage involved artifact detection and correction to enhance the quality and accuracy of the HRV features derived from the IBI data. 
Spectral decomposition for frequency-domain analysis was performed using Welch’s method, providing a reliable estimate of the normalized LF and HF power. 
The normalized power values are computed exclusively from the LF and HF frequency components using the following equations:
\begin{equation}
P_{\text{norm,LF}} = \frac{P_{\text{abs,LF}}}{P_{\text{abs,LF}} + P_{\text{abs,HF}}} \times 100
\end{equation}

\begin{equation}
P_{\text{norm,HF}} = \frac{P_{\text{abs,HF}}}{P_{\text{abs,LF}} + P_{\text{abs,HF}}} \times 100
\end{equation}

Where, $P_{\text{norm,LF}}$ and $P_{\text{norm,HF}}$ represent normalized power in the LF and HF bands, respectively, while $P_{\text{abs,LF}}$ and $P_{\text{abs,HF}}$ denote the absolute power values in these corresponding frequency bands.

\subsubsection{Electrodermal Activity (EDA)}
\label{sec3-4-2}
EDA is a reliable physiological indicator of sympathetic nervous system activity and emotional arousal, reflecting changes in sweat gland activity. The collected EDA signals underwent processing as outlined below.

\textit{Data Cleaning and Artifact Removal:} Initially, a low-pass Butterworth filter with a 10 Hz cutoff frequency was applied to minimize high-frequency noise commonly introduced by environmental interference, sensor movement, and muscle activity. Additionally, a second-order high-pass filter with a Hamming window and a cutoff frequency of 0.05 Hz was used to eliminate low-frequency trends and correct for baseline drift in the signal. After filtering, baseline correction was performed by subtracting the minimum signal value to compensate for variations in the baseline level. Finally, the signal of the second warning was normalized to a range between 0 and 1, ensuring consistency across datasets and facilitating comparability in subsequent analyses.

\textit{Data Analysis:} The processed EDA signals were decomposed into tonic (EDL) and phasic (EDR) components using the cvxEDA algorithm \cite{choi2019feasibility}, facilitating detailed analysis of physiological reactions.
Since the phasic component captures rapid fluctuations in the EDA signal, it serves as the primary indicator of immediate physiological reactions to external stimuli. To maintain consistency across datasets, phasic components were normalized. Given the 4 Hz sampling rate of the EDA signals, data points were grouped into 1-second intervals by averaging every four consecutive values following the warning delivery. 

\subsubsection{Statistical Inference}
\label{sec3-4-3}
To analyze the differences in physiological responses between the Light and Moderate activity conditions, a paired t-test was performed. This analysis aimed to determine whether significant differences existed between the two conditions. Metrics analyzed included mean-HR, std-HR, max-HR, min-NNI, SDNN, RMSSD, NN50, PNN50, normalized LF and HF powers values, and the EDR.

\subsubsection{Features Importance}
\label{sec3-4-4}
HR, HRV and EDR are widely recognized as key physiological indicators of stress, reflecting autonomic nervous system activity and emotional arousal \cite{pop2020electrodermal, ArdecaniShoghli2025, ardecani2024assessing}. 
The feature importance ranking highlights the relative contribution of the HR, HRV, and EDR-related physiological features 
in differentiating stress levels across Light and Moderate activity conditions.

HR and HRV-related features, such as mean-HR, std-HR, max-HR, min-NNI, SDNN, RMSSD, NN50, PNN50, and the normalized power values of LF and HF components, were included in the analysis.  Additionally, the EDR signals for both activity conditions were processed to compute essential statistical features, including mean, standard deviation, median, and peak amplitude as also evidenced by previous research \cite{giakoumis2012using, krantz2004consistency, ushiyama1991physiologic, setz2009discriminating}. These features were selected to capture key physiological characteristics and provide a comprehensive representation of participants' responses to varying activity conditions.

Feature importance was assessed using a Random Forest Classifier, employing the Mean Decrease in Impurity (MDI) method \cite{breiman2001random}, which measures how much each feature contributes to reducing the classification uncertainty across all trees in the ensemble, mathematically expressed as:

\begin{equation}
I(f) = \sum_{t=1}^{T} \sum_{s \in S_t} p_s \cdot \Delta i(s)
\end{equation}

where \( p_s \) represents the proportion of samples reaching node \( s \), and \( \Delta i(s) \) indicates the reduction in impurity (Gini impurity) at that node. Higher importance values indicate that a feature plays a more significant role in classification.

To validate the robustness of our results, we also employed permutation-based feature importance \cite{fisher2019all}, where the values of each feature were randomly shuffled, and the corresponding drop in classification accuracy was recorded.

\subsection{Neurological Data Processing and Analysis}
\label{sec3-5}
In this study we utilized a structured approach for processing EEG data to extract relevant cognitive metrics associated with stress. The methodology follows a data processing framework developed and validated in our previous work \cite{ardecani2025neural}, where EEG signals were preprocessed to mitigate noise and artifacts, filtered to retain relevant neural activity, and analyzed to extract features indicative of cognitive states. In this study, we focus on the critical steps required to extract meaningful stress-related neurological features while ensuring data reliability and consistency. The data processing and analysis involved three key stages: (1) preprocessing, (2) feature extraction, and (3) analysis of alpha and beta band activity for stress assessment.

\subsubsection{EEG Preprocessing}
\label{sec3-5-1}
EEG signals were preprocessed to remove artifacts and enhance signal clarity. Raw EEG data were subjected to removing power line noise artifacts to eliminate 60 Hz electrical interference, applying a 0.1–50 Hz bandpass filter to retain relevant brain activity while filtering out low- and high-frequency noise, using wavelet thresholding to mitigate muscle (EMG), eye movement (EOG), and cardiac (ECG) artifacts, employing MuscIL to detect and remove persistent EMG artifacts, performing segmentation and bad channel interpolation to handle noisy channels exceeding ±150 µV, and re-referencing all channels to the average of the 32 scalp electrodes to standardize the signal. Additionally, Independent Component Analysis (ICA) was applied as a final step to remove residual artifacts. 

\subsubsection{Feature Extraction}
\label{sec3-5-2}
Following preprocessing, feature extraction was performed to obtain spectral power in key frequency bands. We focused on EEG channels in the frontal lobe and prefrontal cortex, as shown in Figure \ref{Cerebral-Cortex-and-physiological}(a), due to their critical role in higher-order cognitive functions such as emotional regulation, stress appraisal, and top-down modulation of the autonomic nervous system during stressful situations \citep{eom2023investigation, lopez2023garments, tsolisou2023eeg, thayer2012meta}. To enhance computational efficiency and eliminate redundancy, Principal Component Analysis (PCA) reduced data dimensionality while preserving significant variance \citep{kastle2021correlation, wang2019detecting}. Given the well-documented relationship between stress and brainwave activity \citep{jun2016eeg, vanhollebeke2022neural, wen2020electroencephalogram, katmah2021review}, we focused specifically on the alpha (8–12 Hz) and beta (12–30 Hz) bands, using Welch’s method to estimate power spectral density (PSD). To account for inter-subject variability, relative power was computed for each frequency band, ensuring comparability across participants.

To investigate the effects of AR warnings, EEG data were subjected to segmentation into 5-second windows before and after each warning event. The pre-warning period served as a baseline reference for resting-state neural activity, while the post-warning period was divided into 125-millisecond intervals to track transient neural responses. 
Our findings \cite{ardecani2025neural} revealed short-lived changes in alpha and beta activity, which returned to ±10\% of baseline levels within 500 milliseconds, aligning with previous studies on stimulus-induced neural oscillations \citep{pfurtscheller1999event, Michelmann2022}. These results confirm that the observed EEG changes were directly induced by the warnings rather than prolonged neural engagement unrelated to the stimuli. 

\subsection{Temporal Relationship Between Cognitive and Physiological Stress Responses}
\label{sec3-5-3}
The stress response involves a coordinated activation of both neurological and physiological systems. At the core of the physiological response are the autonomic nervous system (ANS) and the hypothalamic-pituitary-adrenal (HPA) axis \cite{ulrich2009neural}. When a stressor is perceived, neurons in the hypothalamus stimulate the release of corticotropin-releasing hormone (CRH), which then prompts the secretion of adrenocorticotropic hormone (ACTH) by the pituitary gland. ACTH travels through the bloodstream and activates the adrenal glands, leading to the release of stress hormones such as cortisol, epinephrine, and norepinephrine \cite{kajantie2006effects}. These hormones influence cardiovascular and electrodermal activity, making HR, HRV, and EDR important physiological markers of stress \cite{kim2018stress}.

Understanding the temporal dynamics between cognitive and physiological stress responses is critical, particularly in complex environments where workers must quickly process multi-modal warning signals. Cognitive stress arises as the brain evaluates the relevance, urgency, and potential consequences of external stimuli. This mental stress is reflected in neuro-physiological signals such as Alpha and Beta EEG rhythms, which are sensitive to shifts in attention, alertness, and workload. In contrast, physiological stress responses such as changes in HR, HRV, and EDR often follow cognitive activation and can provide insight into the body’s autonomic regulation under stress.

In our analysis, we explored this temporal relationship by evaluating how changes in EEG-based cognitive stress markers align with or precede physiological responses. For cognitive stress, we focused on the time intervals during which Alpha and Beta power showed their most significant deviations during both light and moderate work activities. For physiological stress, we selected the most informative metric—determined based on feature importance analysis. Pearson’s correlation coefficient was used to assess the strength and timing of associations between these measures, allowing us to investigate whether physiological stress reactions lag behind, align with, or anticipate cognitive stress signals in response to multi-modal warnings.

\section{Results}

\label{sec4}
\subsection{Physiological Stress Responses}
\label{sec4-1}
\subsubsection{Impact of Activity Intensity on HR and HRV Features}
\label{sec4-1-1}
To examine how activity intensity affects physiological stress responses, a paired t-test was conducted to compare HRV physiological features between the Light and Moderate activity conditions. The analysis evaluated  mean-HR, std-HR, max-HR, min-NNI, NN50, PNN50, SDNN, RMSSD and the normalized values of LF and HF components to identify statistically significant differences.

\begin{figure}[H]
    \centering

    \begin{subfigure}{0.9\textwidth}
        \centering
        \includegraphics[width=\textwidth]{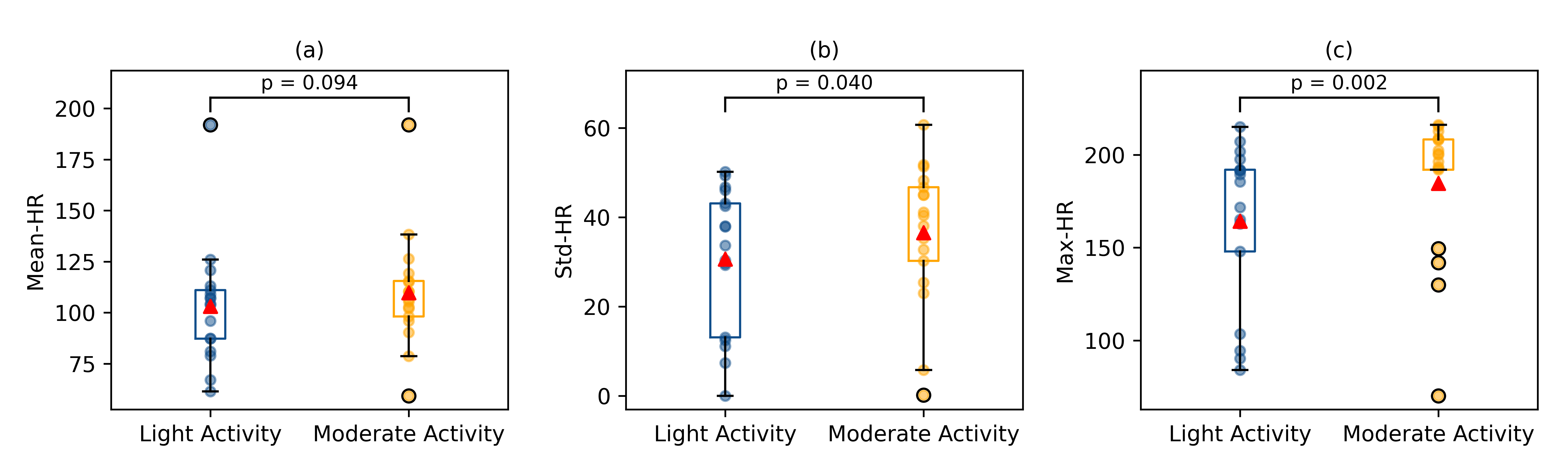}
        \caption{Heart Rate Features}
        \label{fig:hr_stats}
    \end{subfigure}

    \vspace{10pt}

    \begin{subfigure}{0.9\textwidth}
        \centering
        \includegraphics[width=\textwidth]{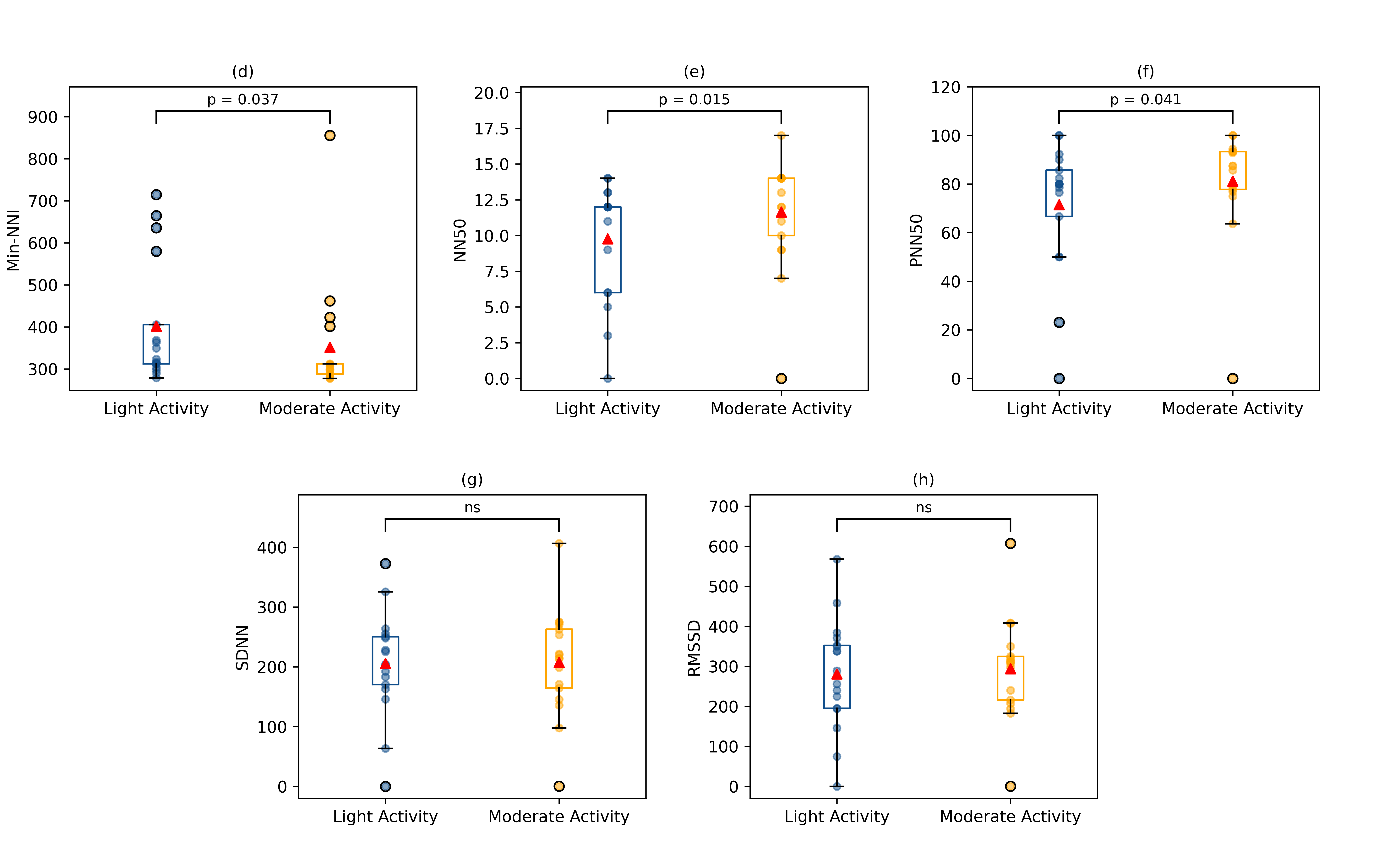}
        \caption{Heart Rate Variability Time-domain Features}
        \label{fig:hrv_time_domain}
    \end{subfigure}

    \vspace{10pt}

    \begin{subfigure}{0.6\textwidth}
        \centering
        \includegraphics[width=\textwidth]{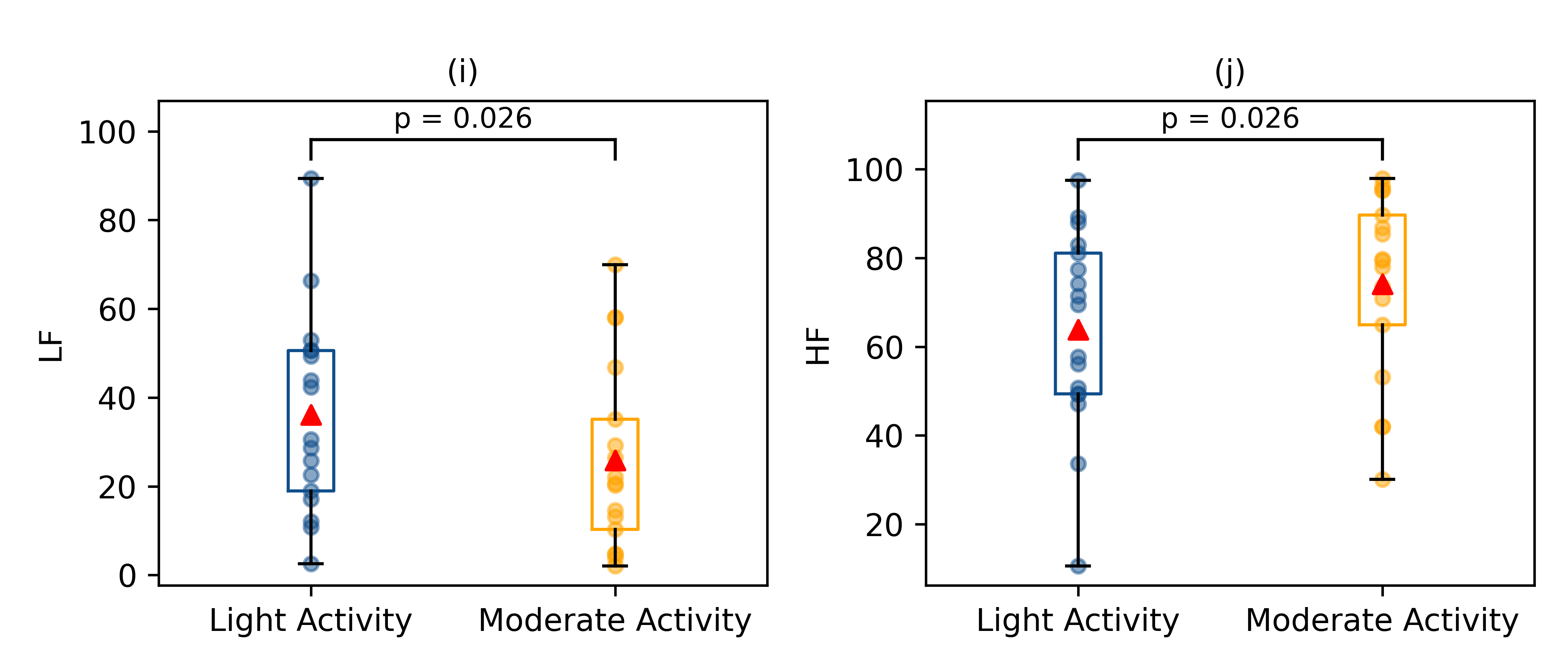}
        \caption{Heart Rate Variability Frequency-domain Features}
        \label{fig:lf_hf}
    \end{subfigure}

    \caption{Comparative analysis of post-warning HR and HRV features between light and moderate activities}
    \label{fig:statistic_HRV}
\end{figure}

Results indicated significant increases in mean-HR (p=0.094), std-HR (p=0.040), and max-HR (p=0.002) during moderate activity, as shown in Figure \ref{fig:statistic_HRV}(a,b,c),
reflecting the cardiovascular system’s adaptive response to heightened exertion levels. Min-NNI was significantly decreased (p=0.037) in the Moderate activity condition, as shown in Figure \ref{fig:statistic_HRV}(d), indicating a reduction in the shortest inter-beat intervals, which reflects an adaptive cardiac response to increased physical effort.  

Parasympathetic indicators, NN50 (p=0.015) and PNN50 (p=0.041), both increased (Figure \ref{fig:statistic_HRV}(e,f)), suggesting altered autonomic activity under increased task demands. However, no statistically significant differences were observed in SDNN and RMSSD between the Light and Moderate activity conditions, as shown in Figure \ref{fig:statistic_HRV}(g,h), indicating stable overall and short-term heart rate variability despite changes in activity intensity.
Furthermore, normalized LF decreased significantly (p=0.026) , whereas normalized HF increased significantly  (p=0.026), suggesting a shift toward sympathetic dominance during moderate activity, as shown in Figure \ref{fig:statistic_HRV}(i,j).

The increase in normalized HF and the corresponding decrease in normalized LF further indicate a shift toward sympathetic dominance, reinforcing the impact of moderate activity on autonomic balance. Additionally, the reductions in min-NNI and increased in parasympathetic indices (NN50 and PNN50) suggest increased cardiovascular and autonomic demands. 
These findings directly address the first objective of the study by demonstrating that increased task intensity triggers changes in HR and HRV-based physiological stress responses. Specifically, the observed shifts in autonomic markers reflect the body's adaptation to physical demands in the presence of AR-enabled warnings, highlighting the impact of AR-based safety interventions across varying task intensities in dynamic occupational settings.
 
\subsubsection{Impact of Activity Intensity on EDR Metrics}
\begin{figure}[H]
    \centering
    \includegraphics[width=0.97\textwidth]{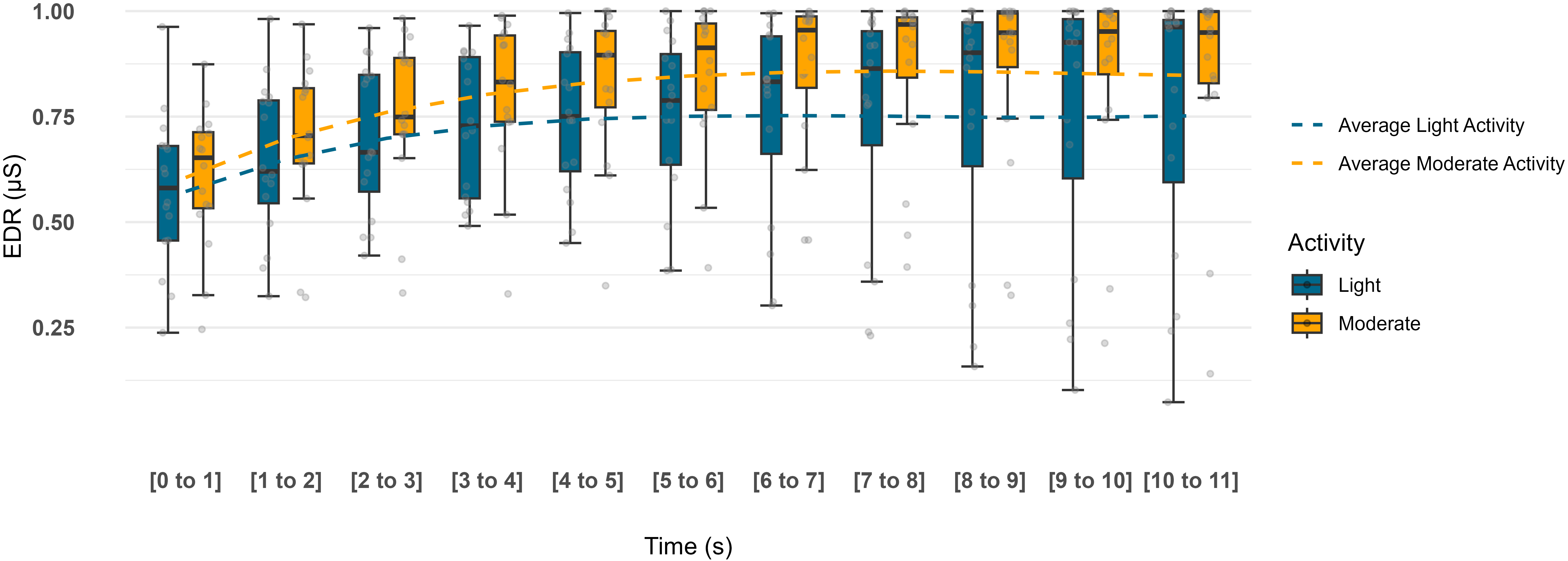}
    \captionsetup{skip=10pt}
    \caption{Comparative analysis of post-warning EDR for light and moderate activities}
    \label{EDR}
\end{figure}

Figure \ref{EDR} illustrates a comparative analysis of the phasic component of Electrodermal Activity (EDA), the Electrodermal Response (EDR), over a 10-second interval following the delivery of the warning for both light and moderate activities. The dashed lines indicate the average EDR across participants for each activity level. As shown, the average EDR during light activity exhibits a steeper initial increase compared to moderate activity, suggesting heightened sensitivity to the warning in less physically demanding tasks. This could be attributed to lower baseline cognitive and physical engagement, making participants more reactive to external stimuli. In contrast, during moderate activity, the physiological response to the warning appears more subdued, possibly due to the cognitive and physical demands of the task requiring greater attentional resources, thereby reducing sensitivity to additional stressors. The box plot further supports this trend, showing that the distribution of participants’ EDR responses is wider in light activity, indicating greater variability in individual reactions. Conversely, the more compact distribution in moderate activity suggests a more uniform physiological response across participants. Light activity stabilizes around the 5th to 6th second, and moderate activity stabilizes around the 6th to 7th second, indicating that while initial responses vary by activity intensity, the recovery trajectory remains relatively uniform across conditions.

The paired t-test results showed a significant increase in EDR (p=0.0003) in the Moderate activity condition, indicating greater autonomic arousal and physiological stress due to higher task demands. These findings highlight that moderate-intensity activity induces measurable physiological changes, particularly in EDR, which signals heightened autonomic arousal and stress response.

\subsection{Features Importance Analysis}
\label{sec4-1-2}

\begin{figure}[H]
    \centering
    \includegraphics[width=0.8\textwidth]{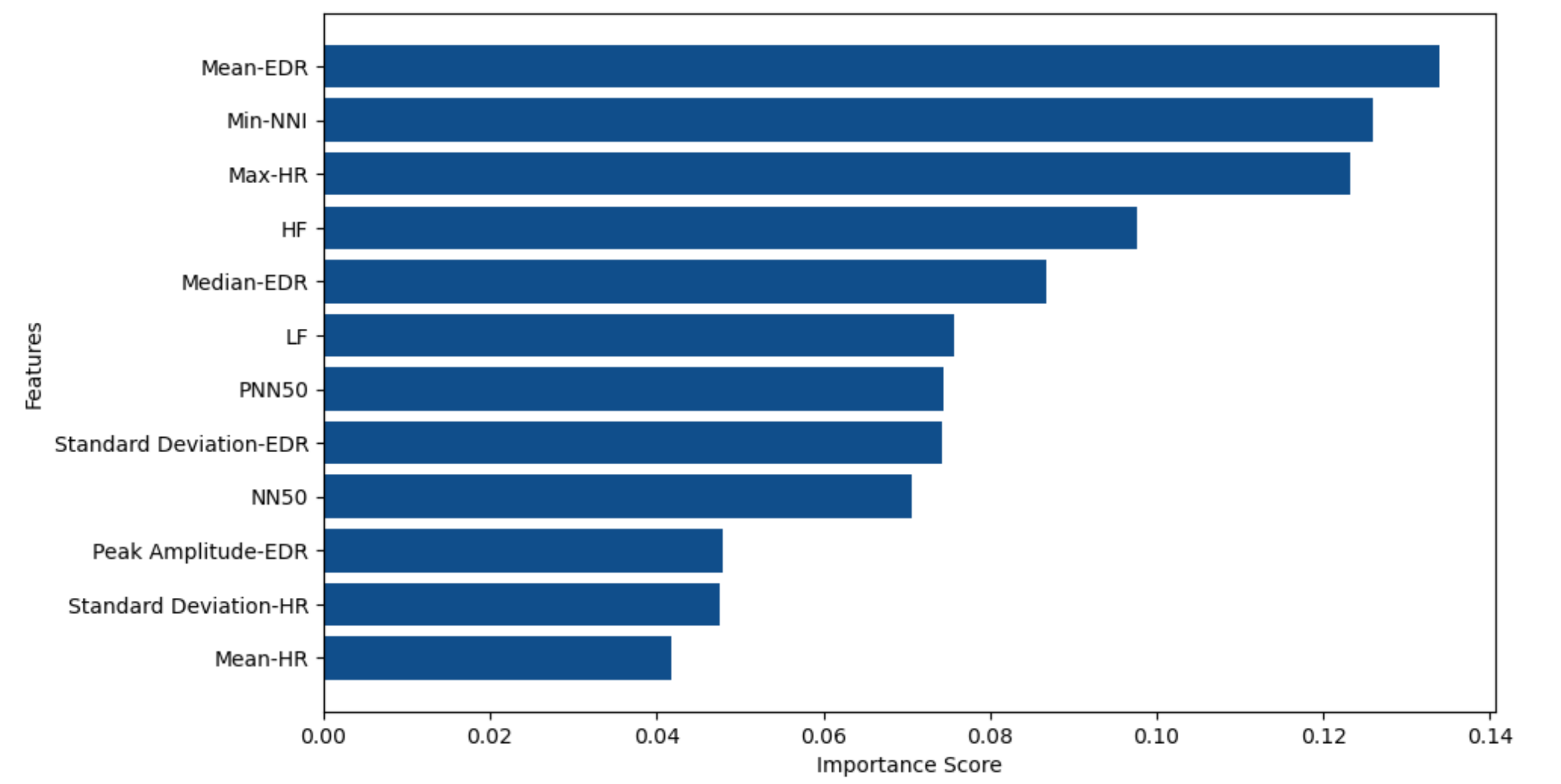} 
    \caption{Feature importance ranking}
    \label{fig:feature_importance}
\end{figure}

Figure \ref{fig:feature_importance} presents feature importance rankings derived from a Random Forest Classifier. 
Mean-EDR emerged as one of the most influential features (importance = 0.134), which reflects sympathetic nervous system arousal and increases under stress, serving as a strong physiological marker for distinguishing between activity levels. Among HRV metrics, Min-NNI (importance = 0.126) and Max-HR (importance = 0.123) were highly influential, highlighting the significance of short-term cardiac dynamics in differentiating activity intensities. Normalized HF (importance = 0.098) and LF (importance = 0.076) components also contributed notably, reinforcing autonomic nervous system regulation as a key factor in stress differentiation. Notably, features such as SDNN and RMSSD were not selected due to the absence of statistically significant differences between Light and Moderate activity conditions.

\subsection{Neurological Stress Responses}
\label{sec4-2}
\begin{figure}[H]
    \centering
    \begin{subfigure}[b]{0.8\textwidth}
        \centering
        \includegraphics[width=\textwidth]{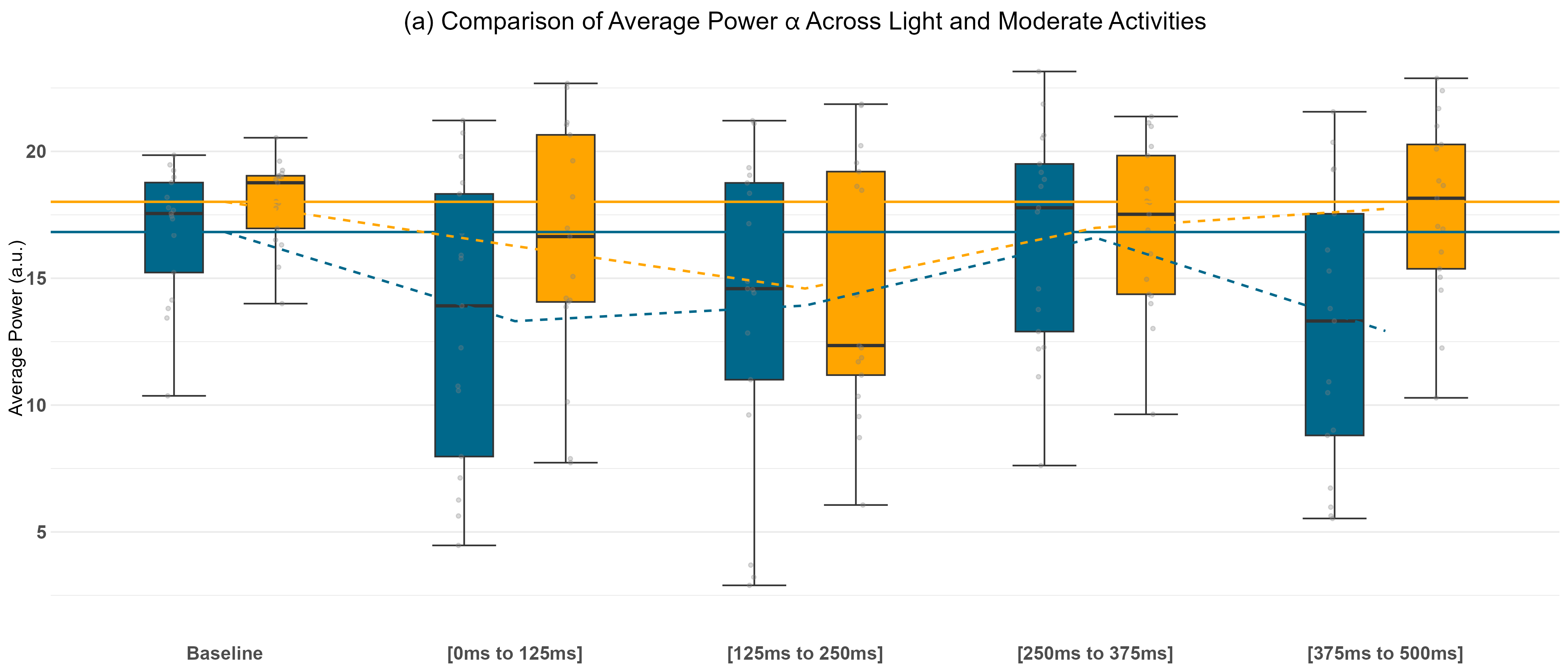}
        \label{fig:alpha}
    \end{subfigure}
    
    \vspace{-1em} 

    \begin{subfigure}[b]{0.8\textwidth}
        \centering
        \includegraphics[width=\textwidth]{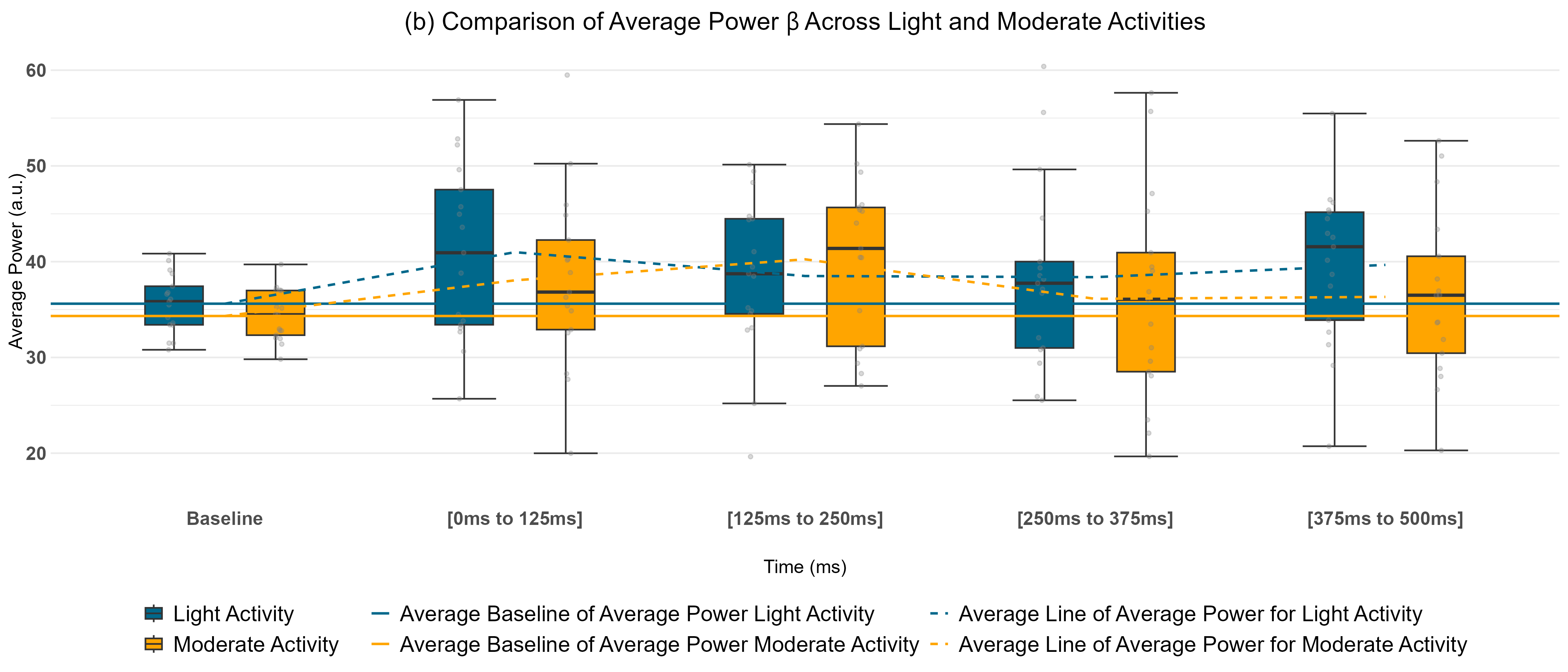}
        \label{fig:beta}
    \end{subfigure}
    \vspace{- 15 pt}
    \caption{Comparison of Alpha ($\alpha$) and Beta ($\beta$) waves in light and moderate activity}
    \label{fig:combined_Stress}
\end{figure}

Figures \ref{fig:combined_Stress}(a,b) illustrate EEG response patterns associated with neurological stress across Light and Moderate activity conditions. Analysis identified distinct patterns in alpha and beta wave activities:
\begin{itemize}
    \item Alpha (${\alpha}$) Power: A consistent decrease in alpha power was observed immediately after the warning in both LA and MA conditions. Notably, the most pronounced reduction in alpha power occurred within the first 125 milliseconds (ms) in the LA condition, whereas in the MA condition, this reduction was observed between 125 and 250 ms across all three warnings. The suppression of alpha power, commonly associated with heightened visual processing and cognitive engagement, was more pronounced in the LA condition, indicating a stronger neural response to stress-related stimuli under lower physical engagement.
    \item Beta ($\beta$) Power: The warning stimulus led to an increase in $\beta$ power, indicating an intensified cognitive response to stress. Notably, beta activity reached its peak within the first 125 milliseconds (ms) in the LA condition, whereas in the MA condition, the peak consistently occurred later, between 125 and 250 ms. Increased beta wave activity reflects elevated cognitive demand and stress. The increase in beta power was more prominent in the LA condition, suggesting that reduced physical exertion enhances neural sensitivity to stress, facilitating a more immediate situational awareness response.
\end{itemize} 

The temporal shift in peak activity for both alpha and beta waves between light and moderate activity conditions demonstrates that neural processing of stress-related stimuli is influenced by physical engagement levels. This modulation of brain responses by task intensity highlights the dynamic interplay between cognitive and physical demands. Such findings underscore the importance of incorporating physical context into the design of AR-based warning systems to ensure that stress levels remain manageable across diverse activity intensities in occupational environments.

\subsection{Temporal Relationship Between Neural and Physiological Stress Responses
}
\label{sec4-3}
To understand the relationship between cognitive stress metrics and physiological responses, we examined their correlation over time. As presented in Section \ref{sec4-2}, our findings reveal that Alpha and Beta waves exhibited their most pronounced changes within the initial 125 ms and the second 125 ms interval for light and moderate activity, respectively. Furthermore, among the physiological indicators considered, EDR demonstrated the highest feature importance, as shown in Section \ref{sec4-1-2}; therefore, EDR was identified as the primary physiological indicator. Given that Alpha power decreases as EDR increases, the strongest negative correlation was identified as the representative correlation. To ensure a consistent basis for comparison, the maximum absolute correlation value was used in this study.

The correlation analysis indicates that Alpha activity during light tasks was most strongly associated with EDR responses at the sixth one-second interval (Correlation = 0.306), while Beta activity peaked in correlation with EDR at the third one-second interval (Correlation = 0.358). In contrast, during moderate tasks, Alpha activity exhibited the strongest correlation with EDR at the third one-second interval (Correlation = 0.266), whereas Beta activity aligned most closely with EDR at the seventh one-second interval (Correlation = 0.395). 

In both activities, Beta activity showed a stronger correlation with EDR than Alpha activity, suggesting that physiological responses are more closely aligned with cognitive stress processes reflected in Beta waves. Furthermore, in light activity, the highest EDR response coincided with the peak correlation of Alpha activity, whereas in moderate activity, the highest EDR response occurred when Beta activity exhibited the strongest correlation. This pattern indicates that Alpha waves, which generally exhibit higher power than Beta waves, show the strongest correlation with EDR during light activity, suggesting a state of moderate cognitive engagement with relatively lower mental effort. In contrast, during moderate activity, Beta waves—typically lower in power but more sensitive to task-related cognitive processing—show the strongest correlation with EDR, indicating that heightened task demands lead to increased cognitive load and a stronger physiological response. This shift in dominant brain wave correlation suggests that as task intensity increases, cognitive effort transitions from a more relaxed, attentional state (Alpha) to an active, task-focused processing mode (Beta), influencing how the body adapts physiologically over time.

These findings underscore the measurable time lag between neural and physiological stress responses, evidenced by the earlier peak correlation of Alpha activity with EDR during light tasks (at sixth one-second interval) compared to the later peak correlation of Beta activity with EDR during moderate tasks (at seventh one-second interval). This temporal progression highlights how task complexity influences the immediacy of cognitive strain and subsequent autonomic regulation. This temporal distinction offers valuable insights into workers’ adaptive mechanisms in multi-modal warning systems and emphasizes the necessity of designing work zone environments that accommodate varying workload conditions, thereby enhancing worker safety and performance in high-risk environments.

\section{Discussion}
\label{sec5}

\subsection{Examining the Stress Indicators of High and Low Neural Responders}

\begin{figure}[H]
    \centering
    
    \begin{subfigure}{0.42\textwidth}
        \centering
        \includegraphics[width=\textwidth]{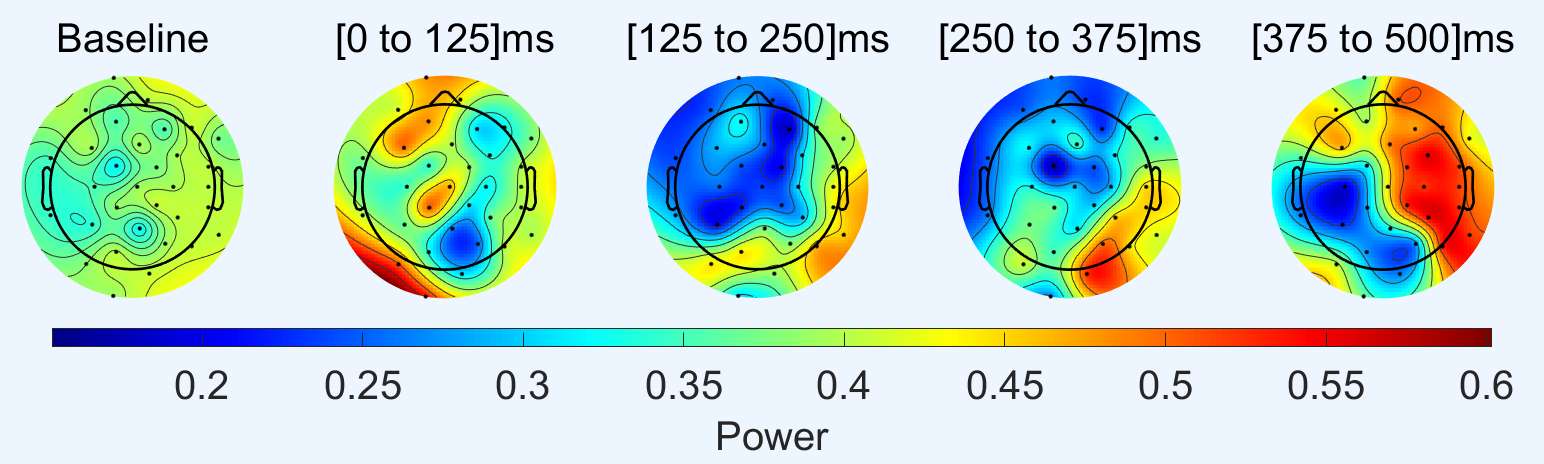}
        \caption{P13 (HNR) Light Activity 
        }
        \label{subfig:P13_beta_L}
    \end{subfigure}
    \hspace{0.02\textwidth}
    \begin{subfigure}{0.42\textwidth}
        \centering
        \includegraphics[width=\textwidth]{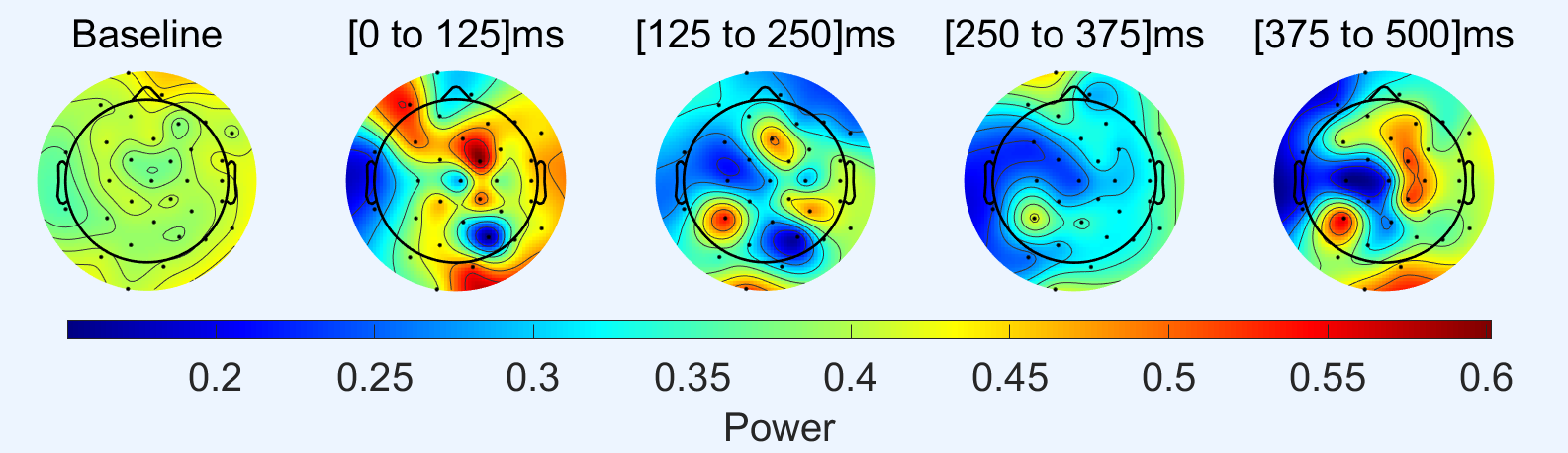}
        \caption{P13 (HNR) Moderate Activity 
        }
        \label{subfig:P13_beta_M}
    \end{subfigure}
    
    \vspace{0.20cm}
    
    \begin{subfigure}{0.42\textwidth}
        \centering
        \includegraphics[width=\textwidth]{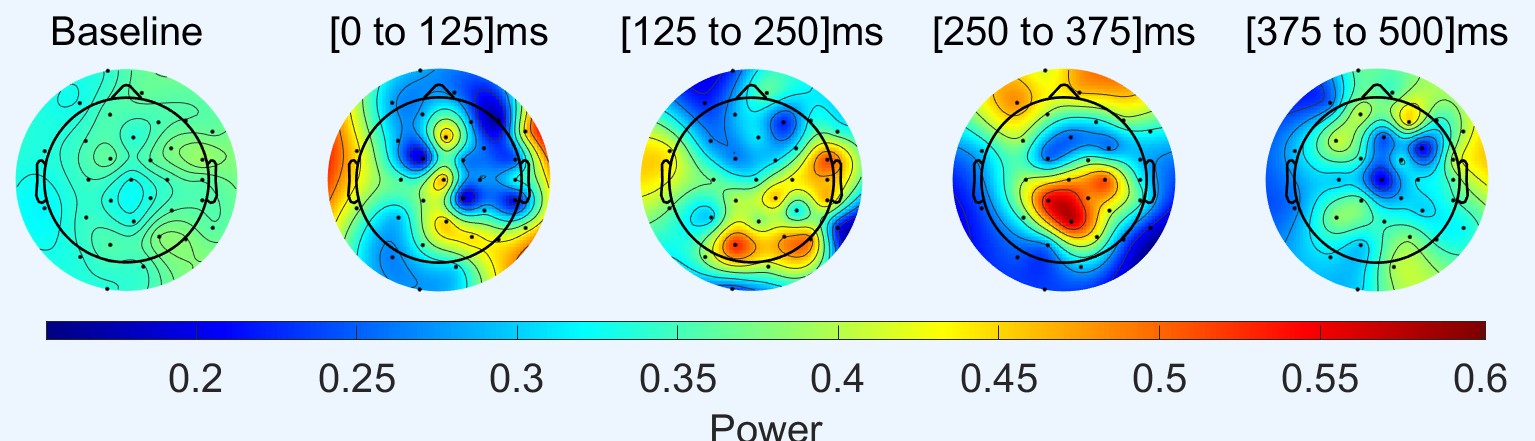}
        \caption{P16 (LNR) Light Activity 
        }
        \label{subfig:P26_beta_L}
    \end{subfigure}
    \hspace{0.02\textwidth}
    \begin{subfigure}{0.42\textwidth}
        \centering
        \includegraphics[width=\textwidth]{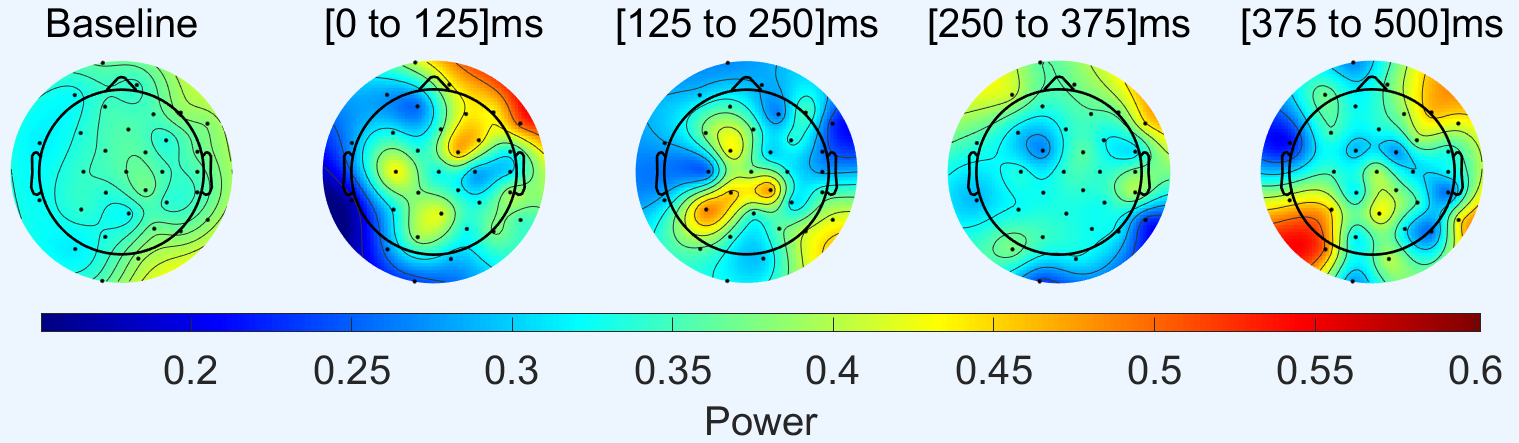}
        \caption{P16 (LNR) Moderate Activity 
        }
        \label{subfig:P26_beta_M}
    \end{subfigure}
    
    \caption{$\beta$ band activity of 
     P13 (HNR) and P16 (LNR) during Light and Moderate Activity}
    \label{fig:P16&13_beta_W2_L_M}
\end{figure}

As highlighted in Section \ref{sec4-3}, beta-band activity demonstrated a stronger correlation with EDR than alpha activity, making beta the primary EEG feature of interest. To explore how changes in beta activity reflect stress responses to AR-enabled safety warnings across light and moderate levels of physical work intensity, we focused on two participants: Participant 13 (P13), a 24-year-old male with three years of construction experience, and Participant 16 (P16), a 24-year-old male with one year of construction experience. Both participants completed tasks under light and moderate intensities. P13 exhibited the greatest change in beta activity between light and moderate tasks, potentially reflecting higher cognitive adaptability to varying work intensities and was thus referred to as the high neural responder (HNR). In contrast, P16 showed minimal changes in beta activity, suggesting a more uniform neural response to post-warning stress across different task intensities and was designated as the low neural responder (LNR). 
As shown in Figure \ref{fig:P16&13_beta_W2_L_M}, P13 exhibits greater beta activity and stronger stress responses to AR-enabled warnings compared to P16 during both light and moderate tasks. Furthermore, this participant showed earlier neural responses within 0–125 ms post-warning during the moderate task, in contrast to P16. The higher level of engagement with the task and responsiveness to AR-warnings, as reflected in beta activity changes, may be partly due to P13’s greater experience in construction environments. However, other intrinsic factors, such as individual anxiety levels, coping strategies, or general cognitive capacity, could also play a role. Conversely, P16’s attenuated neural response may be due to limited field experience or a higher cognitive workload that delayed or suppressed early stress processing.

As shown in Section \ref{sec4-1-2}, the parameters Mean-EDR and Min-NNI have the highest effect in distinguishing between the two activity types. Additionally, as shown in Section \ref{sec4-1-1}, the Min-NNI parameter decreases as activity intensity increases. Among the two selected participants, P13 exhibits a lower reduction in Min-NNI (-52.02) compared to P16 (-61.2) between light and moderate tasks, suggesting that P16 experienced a more pronounced cardiovascular adjustment in response to elevated physical demands. This may reflect reduced physiological resilience under higher task loads, resulting in greater autonomic reactivity during moderate-intensity activity. In contrast, P13’s more moderate change in Min-NNI may indicate heightened autonomic control or greater task engagement, potentially associated with experience-driven physiological adaptability. Furthermore, P13 shows a smaller increase in Mean-EDR (0.155) between light and moderate tasks compared to P16 (0.212), which may reflect a more regulated electrodermal response to rising task intensity. This subtle increase may indicate that, while P13 does respond to the increased workload, his physiological arousal remains more stable—possibly due to greater familiarity with high-stress situations. His experience may contribute to more efficient emotional regulation and better adaption to stress when exposed to AR-enabled safety warnings.

Overall, these findings suggest a potential connection between work experience and other intrinsic factors in shaping how construction workers physiologically respond to hazardous situations. More experienced workers, like P13, may demonstrate more stable autonomic responses, indicating greater resilience and adaptability in high-stress environments. In contrast, less experienced workers may exhibit heightened physiological arousal when exposed to increased task demands. These insights underscore the importance of targeted training to prepare workers for real-world hazards \cite{guegler2021use}. Integrating immersive VR environments with AR-enabled warnings offers a promising approach to safely simulate dangerous scenarios, helping less experienced workers build the necessary cognitive and emotional readiness. This type of training can enhance situational awareness, improve stress regulation, and ultimately contribute to safer and more effective performance in actual work zones \cite{jung2020impact}.

In addition to the effect of work experience, the differing beta activity observed in P16 during moderate-intensity tasks suggests that AR warning systems should be designed to account for task-specific stress dynamics. The contrast between P16 and P13 highlights how individual differences can influence physiological responses to task demands, emphasizing the value of personalized task assignments that account for each worker's capacity to handle cognitive and physical stressors. Research by Ojha et al. (2024) \cite{ojha2024quantifying} has suggested that biometric sensing can optimize task assignments to enhance worker safety and efficiency.
Furthermore, in hazardous environments, identifying reliable and interpretable psychophysiological indicators is crucial for enhancing the effectiveness of real-time safety interventions. Prior research has also highlighted the importance of real-time physiological monitoring in managing risks on construction sites. For instance, Jebelli et al. (2019) \cite{jebelli2019applicationI} demonstrated that early detection of cognitive overload through biosensing can improve the timing of interventions and reduce the likelihood of human error. While cognitive overload has received significant attention, growing evidence suggests that stress may be an even more critical factor influencing worker performance and safety outcomes in high-risk environments \cite{TuerxunWaili_2020, pei2024identifying}. Our findings show that the Mean-EDR marker, as demonstrated through the response of Participant 13 and supported by feature importance analysis, reflects the combined effects of task intensity and work experience. This physiological indicator can provide meaningful, real-time insights into individual stress responses during high-demand tasks, enabling more responsive and adaptive safety interventions. Building on this, the integration of AR systems with real-time stress detection presents a novel opportunity: by identifying early autonomic stress responses, such systems could deliver personalized, just-in-time interventions. This approach could extend beyond construction, offering value in other high-stakes environments such as aviation, manufacturing, and defense operations where human error under stress can have critical consequences.

\subsection{Insights from Neuro-physiological Stress Responses to AR Warnings}
Our findings align with the growing body of literature that identifies HRV and EDR as reliable indicators of autonomic stress. Specifically, the observed increase in mean-HR, max-HR, coupled with decreases in min-NNI under the moderate-intensity task align with the findings of Kim et al. \cite{kim2018stress}, who reported that heightened physical workload leads to reduced HRV and increased sympathetic nervous system activation. Similarly, our results demonstrate a significant increase in EDR following AR warnings during moderate-intensity tasks, supporting Pop-Jordanova et al. \cite{pop2020electrodermal}, showing that skin conductance levels are a reliable measure of stress responses in high-risk environments.
Beyond these autonomic indicators, our EEG analysis reveals consistent results with previous studies. In particular, alpha power suppression and concurrent beta power elevation after AR warnings, were consistent with evidence that alpha desynchronization occurs during high-stress conditions as the brain reallocates resources for heightened situational awareness \cite{8758154}, while increased beta power is associated with increased cognitive engagement and stress \cite{vanhollebeke2022neural}. 

From a theoretical standpoint, these neural shifts reflect the reallocation of attentional resources under stress, an idea that aligns with the resource theory of attention \cite{kahneman1973attention}, which suggests that cognitive resources become taxed under dual demands of physical and mental workload. Our findings show that while alpha and beta power peak earlier in light-intensity tasks, they exhibit a delayed but pronounced increase under moderate-intensity conditions. This latency suggests that when physical exertion rises, the brain requires additional time to redirect resources to process AR stimuli, consistent with Cohen et al. (2014), who demonstrated that cognitive load distribution depends on task complexity and physical strain \cite{cohen2014analyzing}.

From a practical standpoint, these findings underscore the importance of designing AR safety systems that balance enhanced awareness with the risk of cognitive overload and stress. One practical strategy involves adapting the timing or complexity of AR warnings based on real-time physiological cues. For example, if elevated EDR or a drop in HRV is detected, the system could momentarily simplify new information, allowing workers to regain cognitive equilibrium and focus on time-sensitive warnings. Such adaptive interfaces hold promise for optimizing worker safety and performance without unduly increasing stress.

\subsection{Relationship Between Neural and Physiological Stress Responses}
The brain and heart are connected through a dynamic, bidirectional communication system that involves both autonomic regulation and hemodynamic signaling. This connection enables the brain to modulate cardiac activity while simultaneously receiving feedback from physiological states. 
Electrodermal responses (EDR), driven by sympathetic nervous system activation, offer additional insight into this brain–body interaction by capturing subtle arousal and stress patterns. Within our study's context, these interlinked systems reveal how neural activity, as measured by EEG, and physiological stress indicators, as measured by HR, HRV, and EDR, are temporally connected and collectively reflect workers' stress responses to AR over time.

The correlation analysis showed that alpha activity exhibited the strongest relationship with EDR during light-intensity tasks, while beta activity was more strongly correlated with EDR in moderate-intensity tasks. This finding is consistent with Attar et al.  \cite{attar2021stress}, who reported that EEG and ANS-related markers exhibit activity-dependent correlations, with higher cognitive workload leading to increased synchronization between EEG beta power and autonomic responses. Similarly, other studies \cite{hayashi2009beta, arsalan2022human} have shown that beta-band activity becomes more prominent during states of high strain or heightened alertness, reinforcing its role as a key indicator of cognitive stress in demanding environments.

We observed a time lag between EEG and physiological responses of approximately 4875 ms and 5750 ms for light and moderate activity, respectively. Neural changes occurred within the first 0–125 ms and 125–250 ms post-warning for light and moderate activity, respectively, while peak correlations with EDR emerged at the sixth and seventh one-second intervals. This temporal gap, typically spanning the fourth to sixth one-second intervals, suggests that neural responses precede autonomic stress adaptations.
This aligns with Cohen et al. \cite{cohen2014analyzing}, who found that pre-stimulus neural activity influences subsequent autonomic responses. Our results support the hypothesis that EEG markers provide an early indication of stress, while physiological responses such as HR, HRV and EDR reflect a delayed but sustained autonomic reaction.

Our results differ from Jebelli et al. \cite{jebelli2018eeg}, who reported that EEG stress responses in construction workers were more variable and less consistent across individuals. The discrepancy may be due to differences in experimental design, as our study utilized a within-subject framework with baseline normalization, reducing inter-individual variability. This suggests that incorporating both EEG and physiological markers in stress assessment may enhance measurement reliability compared to using EEG alone.

\subsection{Practical Implications}
Beyond enhancing our understanding of stress responses, this research provides a pathway for monitoring of workers’ cognitive and physiological states in high-risk settings such as roadway work zones. By leveraging psycho- and nuero- physiological data (e.g., HR, HRV, EDR, EEG), the proposed framework enables adaptive safety systems that dynamically adjust warning intensity, timing, or modality based on each worker’s stress level. For example, if wearable sensors detect rising electrodermal responses, the system could reduce the frequency of non-critical alerts, prioritize the most hazardous threats, or escalate interventions only when stress surges threaten performance and safety.

Additionally, personalizing AR warnings in real time can significantly improve risk awareness, particularly for less experienced workers. This customization might include delivering more prominent visual cues to novices or providing targeted training overlays that help them manage stress and develop resilience before they encounter real roadway hazards. Alongside advanced VR and AR-based training modules, these insights can lead to more effective skill-building programs, allowing workers to practice high-demand tasks in controlled yet realistic environments. Over time, trainees can learn to regulate their own stress responses, carry out tasks more safely, and develop confidence in reacting to complex on-site situations.

Further, implementing multi-sensor integration into existing project management and safety oversight platforms can give supervisors a dashboard-style view of workforce stress, enabling data-driven decisions regarding task rotation, break schedules, and assignment of specialized jobs to individuals best suited to manage the associated cognitive load. Such approaches can reduce accidents and costly downtime stemming from human error or fatigue-related incidents. In turn, this data-informed approach to workforce management supports a more resilient, attentive, and productive labor force, benefitting both worker well-being and project efficiency.

Finally, the potential of neurophysiology-informed technology extends beyond construction. Industries such as manufacturing, mining, aviation, and defense face similarly demanding conditions where human error under stress can be catastrophic. By tailoring the AR and sensing framework to each domain’s unique hazards and workflow, the same principles of real-time stress detection, adaptive intervention, and personalized training can promote safety across a broad spectrum of high-stakes environments.

\section{Conclusion}
\label{sec6}
This study focused on the analysis of workers' neuro-physiological stress responses to multi-sensory AR warnings under light and moderate task intensities. By capturing HRV, EDR, and EEG signals within a high-fidelity virtual reality environment of a roadway work zone, we demonstrated that AR-based interventions elicit distinct patterns of autonomic nervous system (ANS) regulation, sympathetic nervous system (SNS) activation, and brain activity across different task intensities. 

HR and HRV metrics revealed notable shifts in cardiac autonomic regulation, reflecting heightened physiological demands during moderate-intensity tasks and a shift toward sympathetic dominance as task intensity increased. Similarly, EDR analysis demonstrated a significant increase in physiological arousal during moderate activity. This suggests that higher task demands induce greater autonomic stress responses over time and reduce initial sensitivity to external stimuli, highlighting the importance of evaluating designed multi-modal AR warnings tailored to task intensity.
EEG analysis indicated heightened neural arousal and cognitive stress following AR warnings. Variations in neural response timing between light and moderate activity conditions suggest that task intensity influences stress processing, underscoring the need for adaptive AR systems that align with workload demands.
Among the evaluated features of HR, HRV, and EDR, Mean-EDR was identified as the most influential, highlighting its strength in capturing sympathetic arousal and distinguishing stress responses across activity conditions. Among HRV-related features, Min-NNI, exhibited the highest contributions, underscoring the importance of short-term heart rate fluctuations in capturing physiological responses to varying exertion levels. Furthermore, correlation analysis between alpha and beta activity and EDR revealed distinct psychophysiological response patterns influenced by task intensity following multi-modal AR warnings. Notably, beta activity—a neural marker associated with heightened arousal and alertness—showed stronger correlations with EDR, highlighting its dominant role in reflecting sympathetic arousal and cognitive stress responses driven by differences in physical work intensity in the post-AR warning period. These findings support the potential for integrating EEG-EDA coupling in future adaptive safety systems to better monitor and manage stress in high-intensity work environments.

Our findings offer three main contributions. First, the study underscores the merit of the develoepd multimodal human-sensing approach, linking autonomic (HRV, EDR) and neural (EEG) signals to better capture the timing and magnitude of stress. Second, the results show how stress responses vary with task intensity: alpha activity tends to peak earlier and fade faster under lighter loads, while beta activity becomes more pronounced under moderate conditions, suggesting differing cognitive resource allocations. Third, we demonstrate that the interplay between neural and physiological responses can inform the design of adaptive AR safety systems. For instance, monitoring increases in beta power, along with heightened EDR, could guide real-time adjustments to warning intensity or timing, minimizing the risk of cognitive overload.

A few limitations of this study can guide future research. While the virtual reality simulation provided a controlled and low-risk setting for experimentation, it may not entirely replicate the dynamic and unpredictable nature of real-world work zones. Additionally, although we analyzed stress responses during two task intensities, our experimental framework may not encompass the complete spectrum of physical and cognitive demands that workers face in real-world scenarios. 
Additionally, this study did not examine individual differences such as age, prior exposure to AR technologies, cognitive styles, and disabilities that may impact how workers perceive and respond to AR warnings. Future research may explore these factors to understand the effectiveness of AR across a diverse workforce.

\section{Acknowledgments}
This research was supported in part by the National Science Foundation under Award Number 1932524. The authors would like to acknowledge all participants in this study for their time and contributions. We also acknowledge the efforts of Sepehr Sabeti in developing the Virtual Reality model of the work zones and assisting with data collection, as well as Amit Kumar for his contributions to data collection.

\bibliographystyle{elsarticle-num-names} 
\bibliography{Refs}

\end{document}